\documentclass[preprint,pre,preprintnumbers,superscriptaddress]{revtex4-1}
\usepackage{latexsym}
\usepackage{amsmath}
\usepackage{graphicx}
\usepackage{bm}
\usepackage{epsf}
\usepackage{epsfig}
\usepackage{dcolumn}
\usepackage{color}

\begin{document}

\title{Control of Surface Wettability via Strain Engineering}
\author{Wei Xiong}
\affiliation{Department of Engineering Mechanics and Center for Nano and Micro Mechanics, Tsinghua University, Beijing 100084, China}
\author{Jefferson Zhe Liu}
\email{zhe.liu@monash.edu}
\affiliation{Department of Mechanical and Aerospace Engineering, Monash University, Clayton, VIC 3800, Australia}
\author{Zhiliang Zhang}
\affiliation{Department of Structural Engineering, Norwegian University of Science and Technology (NTNU), N-7491 Trondheim, Norway}
\author{Quanshui Zheng}
\email{zhengqs@tsinghua.edu.cn}
\affiliation{Department of Engineering Mechanics and Center for Nano and Micro Mechanics, Tsinghua University, Beijing 100084, China}

\begin{abstract}
Reversible control of surface wettability has wide applications in lab-on-chip systems, tunable optical lenses, and microfluidic tools. Using a graphene sheet as a sample material and molecular dynamic (MD) simulations, we demonstrate that strain engineering can serve as an effective way to control the surface wettability. The contact angles $\theta$ of water droplets on a graphene vary from 72.5$^\circ$ to 106$^\circ$ under biaxial strains ranging from $-10\%$ to $10\%$ that are applied on the graphene layer. For an intrinsic hydrophilic surface (at zero strain), the variation of $\theta$ upon the applied strains is more sensitive, i.e., from 0$^\circ$ to 74.8$^\circ$. Overall the cosines of the contact angles exhibit a linear relation with respect to the strains. In light of the inherent dependence of the contact angle on liquid-solid interfacial energy, we develop an analytic model to show the $\cos \theta$ as a linear function of the adsorption energy $E_{ads}$ of a single water molecule over the substrate surface. This model agrees with our MD results very well. Together with the linear dependence of $E_{ads}$ on biaxial strains, we can thus understand the effect of strains on the surface wettability. Thanks to the ease of reversibly applying mechanical strains in micro/nano-electromechanical systems (MEMS/NEMS), we believe that strain engineering can be a promising means to achieve the reversibly control of surface wettability.
\end{abstract}

\pacs{Valid PACS appear here}
\maketitle

\section{Introduction}

Understanding the surface characteristics and controlling the wettability of solid surfaces are fundamental topics in chemical physics and serves as a basis in many applications \cite{Nakajima2011}. The Lotus-Leaf-like super-hydrophobic surface, for example, has attracted enormous attention in the past decades because of the superior water-repellent and self-cleaning properties \cite{Quere2008}. On the other hand, super-hydrophilicity of surfaces has also seen itself wide applications, e.g., improving the filtration efficiency of polymer filter thin films \cite{Sun2005} and the boiling heat transfer efficiency in heat pipes \cite{Li2008, Chen2009}. Recently active and reversible control of the surface wettability is becoming a very attractive research topic and various applications are proposed and demonstrated.

Engineering the solid surface structures is the most widely used method to control the wettability, e.g., via introduction of surface roughness/defects and changes of the chemical properties \cite{Feng2002, Sun2005, Quere2008, Zheng2010}. It is reported that the surface energy of graphene increased as more defects were induced, leading to a hydrophilic nature \cite{Shin2010}. Rafiee et al. found that the contact angle of graphene film can be controlled when the graphene film was dispersed on a substrate by performing high-power ultra-sonication in the solvent with a controlled proportion of acetone and water \cite{Rafiee2010}. Despite the great success of this approach, there are several trade-offs. Inducing roughness, defects and chemical groups will affect the structural integrity and thus reduce the robustness of the performance in practice (e.g., under harsh environments) \cite{Callies2005}, together with some other undesired consequences, e.g., enhanced slip friction force for the motion of fluid droplets, and so on. Additionally, most of the methods to change the solid surface structure and wettability are usually permanent and lack of active controllability and reversibility, which are highly desired in novel micro/nano-fluidic devices \cite{Krupenkin2004, Callies2005}.

Reversible controls of wettability have been recently demonstrated by some elegant methods, including the light irradiation \cite{Ichimura2000, Feng2001}, the electrochemical surface modifications \cite{Abbott1995, Gallardo1999}, applying an electric field \cite{Lahann2003}, and the thermal treatment \cite{Mettu2008}. But it still suffers some limitations, e.g., a small controllable range of contact angles (about 11$^\circ$ \cite{Feng2001, Sun2005}), a large hysteresis loop \cite{Ichimura2000, Lahann2003}, delayed dynamic motion of liquid droplets \cite{Ichimura2000}, and a low number of life cycles \cite{Feng2001}. Most of these limitations can be attributed to the fact that the wettability changes are accomplished by the conformational transition of the surface molecular structures.

In this paper, we will demonstrate that the strain engineering can serve as an effective way to reversely control the wettability of an atomically smooth surface without damaging the structure. Graphene is selected as the solid substrate, because of its superior mechanical, electronic, and bio-compatibility properties, which render it an ideal material building-block in the nano-fluidic devices \cite{Mugele2012}. We studied the wetting behaviour of water droplets on graphene sheets with a biaxial strain $\epsilon$ range from $-10\%$ to $10\%$ using molecular dynamics (MD) simulations. The contact angles $\theta$ of water droplets on the graphene surface vary from 72.5$^\circ$ to 106$^\circ$ in this strain range. For the intrinsic hydrophilic surfaces (i.e., at zero strain), the variation of $\theta$ upon the applied strains is even more sensitive than that of intrinsic hydrophobic surfaces (e.g., from 0$^\circ$ to 74.8$^\circ$). Overall we find a linear relationship between the $\cos \theta$ and $\epsilon$. In the end, an analytical model will be presented to explain this observation.

\section{Simulations and Methodology}

\begin{figure*}[htbp]
\centerline{\includegraphics[width=0.6\textwidth]{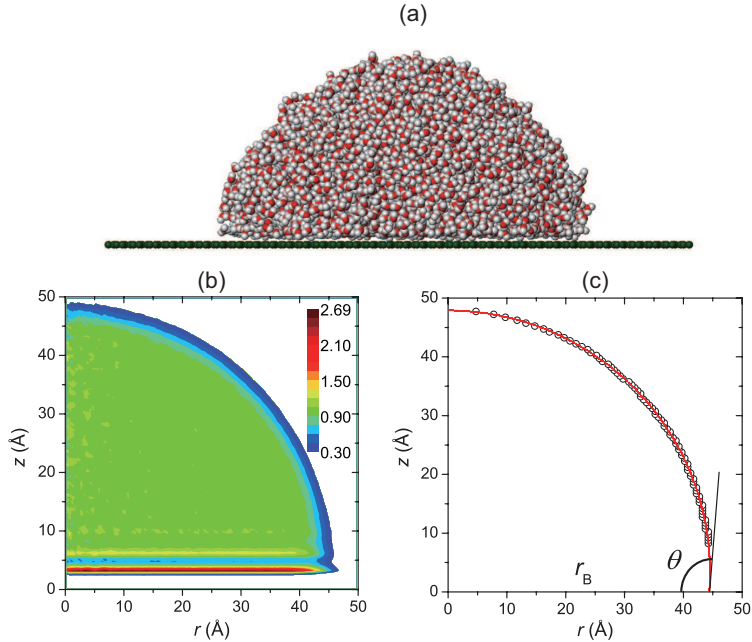}}
\caption{(Color online) (a) A water droplet on strained graphene where the carbon atoms are fixed; (b) Density map of a water droplet with 6600 water molecules on graphene sheet at $0\%$ strain, which has been averaged along the radial direction; (c) Determination of the contact angle $\theta$ and the base curvature $1/r_B$ by fitting a circle to the free surface of the water droplet in (b).}
\label{fig1}
\end{figure*}

Our molecular system is illustrated in Fig. \ref{fig1}(a): a water droplet on a single graphene sheet. The graphene sheet is biaxially strained, ranging from $-10\%$ to $10\%$. Accordingly the carbon-carbon bond length $a_{\text{CC}}$ changes from $0.9a_{\text{CC}_0}$ to $1.1a_{\text{CC}_0}$, where $a_{\text{CC}_0}$ denotes the carbon-carbon bond length of strain-free graphene (e.g., $a_{\text{CC}_0} = 0.142$ nm). We placed four different sizes of water droplets on the strained graphene sheets, i.e., 748, 1885, 3709, 6600 water molecules, respectively. In our MD simulations, the positions of carbon atoms are fixed. We have tried flexible graphene substrates with applied $0\%$ and $10\%$ biaxial strain and found a less than 5$^\circ$ difference in the calculated contact angles from those of the corresponding rigid substrates.


All MD simulations were performed with the LAMMPS code \cite{Plimpton1995}. A time step of 1.0 fs was used and the total simulations time was about a few nanoseconds. We used the CHARMM force field and the SPC/E model \cite{Berendsen1987} for water with the SHAKE algorithm \cite{Ryckaert1977}. The van der Waals (vdw)  interactions between water molecules and carbon atoms were described by a Lennard-Jones (LJ) potential between oxygen and carbon atoms, i.e. $\phi (r)=4\epsilon ((\sigma /r)^{12} - (\sigma /r)^{6} )$. The vdw forces were truncated at 1.2 nm with long-range Columbic interactions computed using the particle-particle particle-mesh (PPPM) algorithm \cite{Hockney1992}. Water molecules were kept at a constant temperature of 300 K using the Nos\'e-Hoover thermostat.

While maintain the same hexagonal lattice with graphene for the monolayer solid substrates, we select three sets of LJ parameter for the water-solid interactions, corresponding to macroscopic contact angles of 91.2$^\circ$, 52.1$^\circ$, and 133$^\circ$ on the zero strained solid surfaces. There are two motivations. First, we can study the strain engineering effects on the contact angles for substrates with different wettability. Second, the contact angle of graphite measured in experiments is scattered, varying from 0$^\circ$ to over 115$^\circ$. It should be noted that value of  about 90$^\circ$ is commonly accepted \cite{Werder2003, Mattia2008}. The parameters of $\sigma = 0.3190$ nm and $\epsilon = 4.063$ meV were adopted to reproduce such an angle \cite{Werder2003}. We kept the same $\sigma$ value but altered the parameter $\epsilon$ as 5.848 meV or 1.949 meV to represent the hydrophilic or hydrophobic strain-free solid substrates, respectively \cite{Werder2003}. We label these substrates as graphene, hydrophilic-surface and hydrophobic-surface, accordingly.

It usually took a few hundred picoseconds to reach equilibrium and the simulations were then continued for two more nanoseconds to collect data. From the MD molecular trajectories, water density maps were obtained by introducing a 3-dimensional grid with the size of each cell as $0.5 \text{\AA} \times 0.5 \text{\AA} \times 0.5 \text{\AA}$. Reducing the 3D mesh into a 2-dimensional density map by averaging along radial direction leads to Fig. \ref{fig1}(b), which is the density map of the droplet with 6600 water molecules on a graphene sheet with $0\%$ strain. We further averaged the 2D density maps of 10000 frames of the trajectory in a total of 2 ns duration. To obtain the water contact angle from such a map, a two-step procedure was adopted following the reference \cite{Werder2003}. First, the boundary of the droplet surface was determined within every single layer that was vertical to $z$ direction by using 0.5 $\text{g/cm}^3$ as a critical density \cite{Weijs2011}. Second, a circular best fit through these points was extrapolated to the solid surface where the contact angle $\theta$ was measured as shown in Fig. \ref{fig1}(c). Note that the points of the surface below a height of 8 \AA\ from the solid surface were not taken into account for the fit, to avoid the influence from density fluctuations at the liquid-solid interface.

\section{Results and Discussions}

\begin{figure*}[htbp]
\centerline{\includegraphics[width=0.8\textwidth]{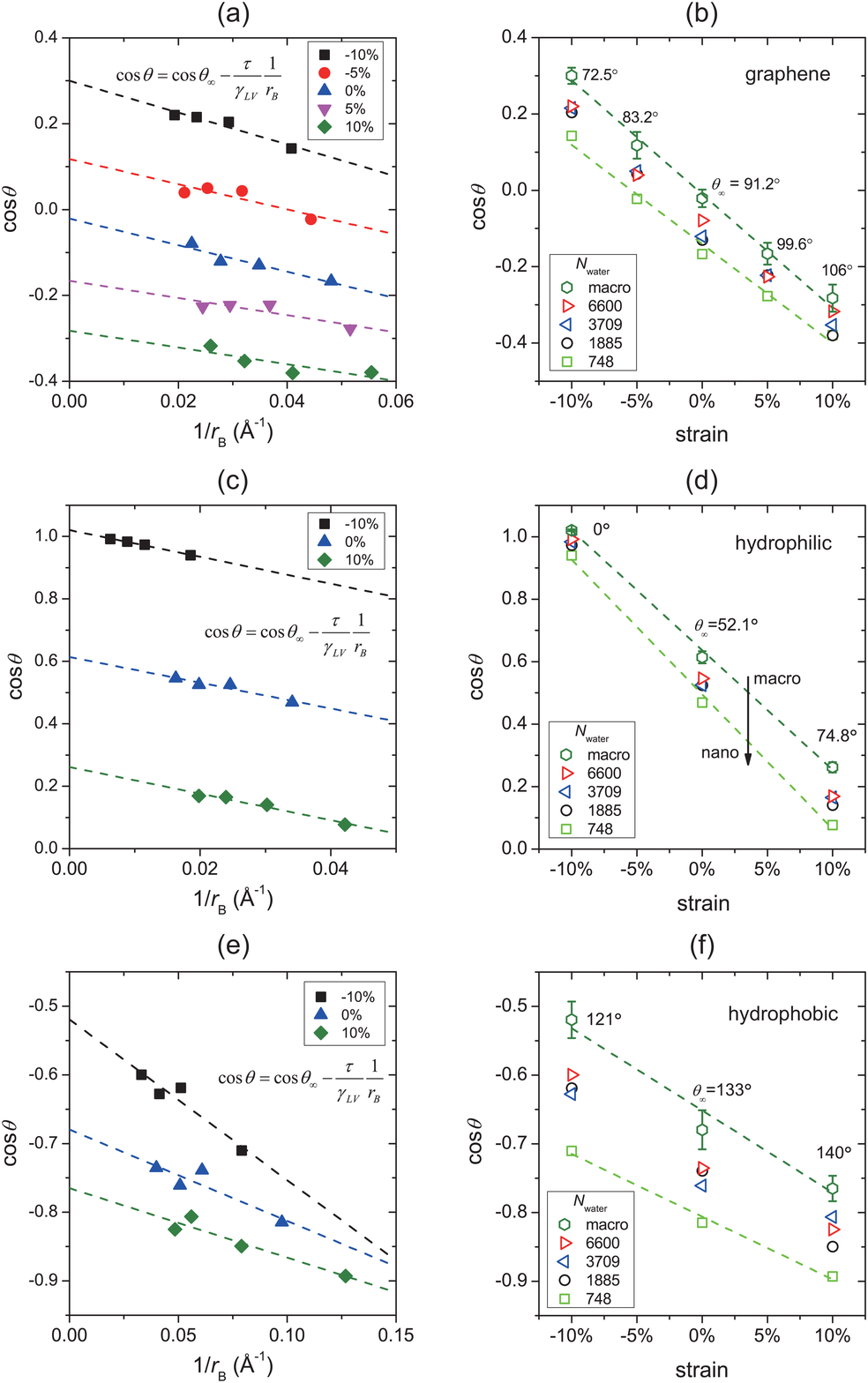}}
\caption{(color online) Cosine of the contact angle $\theta$ as a function of the base curvature $1/r_B$ of the droplets on (a) graphene, (c) hydrophilic-surface and (e) hydrophobic-surface; Cosine of the contact angle $\theta$ as a function of the biaxial strains applied on (b) graphene, (d) hydrophilic-surface, (f) hydrophobic-surface.}
\label{fig2}
\end{figure*}

Fig. \ref{fig2} summarized the contact angles of the water droplets with different sizes on the strained sheets of graphene (Fig. \ref{fig2}(a-b)), hydrophilic-surface (Fig. \ref{fig2}(c-d)), and hydrophobic-surface (Fig. \ref{fig2}(e-f)). The measured microscopic contact angles $\theta$ depend on the size of the droplets. According to the modified Young's equation \cite{Werder2003, Amirfazli2004}, the macroscopic contact angle $\theta_\infty$ can be related to the microscopic contact angles as:
\begin{equation}\label{eq1}
    \cos\theta = \cos\theta_\infty - \frac{\tau}{\gamma_{LV}}\frac{1}{r_B}
\end{equation}
where $\tau$ is the line tension, $1/r_B$ is the fitted base curvature of the water droplets, and $\gamma_{LV}$ is the interfacial free energy per unit area for the liquid-vapor interface. In Fig. \ref{fig2}, the cosines of contact angles measured in our MD simulations follow the linear relation very well with respect to $1/r_B$ (Eq.(\ref{eq1})). Extrapolating the linear relations in Fig. \ref{fig2}(a,c,e) yields the macroscopic contact angles $\theta_{\infty}$.

Fig. \ref{fig2}(b), (d), and (f) depict the contact angles $\theta$ and $\theta_{\infty}$ (hexagon symbols) as a function of the applied strains. The cosines of the (macroscopic) contact angles exhibit a linear relation with the mechanical strains. For the three types of surfaces, positive strains (stretching) always result in a decrease of $\cos \theta$ and thus the increase of contact angle $\theta$, and vice versa. The magnitude of the angle changes is, however, quite different for the three types of solid surfaces. The $\theta_{\infty}$ of graphene varies from 72.5$^\circ$ to 106$^\circ$ for the applied strain from $-10\%$ to $10\%$. On the hydrophobic-surface, the $\theta_{\infty}$ varies from 121$^\circ$ to 140$^\circ$. For the hydrophilic-surface, the change is much more significant, from 0$^\circ$ to 74.8$^\circ$. In comparison, other methods have a smaller controllable range of contact angles. For example, by using an electric field, the (receding) contact angles of water droplets on a low density monolayer can be reversibly controlled between 20$^\circ$ and 50$^\circ$ \cite{Lahann2003}. Subject to a light irradiation, the largest contact angle change on langmuir-blodgett films with photoresponsive fluorine-containing azobenzene polymer is about 11$^\circ$ \cite{Feng2001}.

\begin{figure*}[htbp]
\centerline{\includegraphics[width=0.6\textwidth]{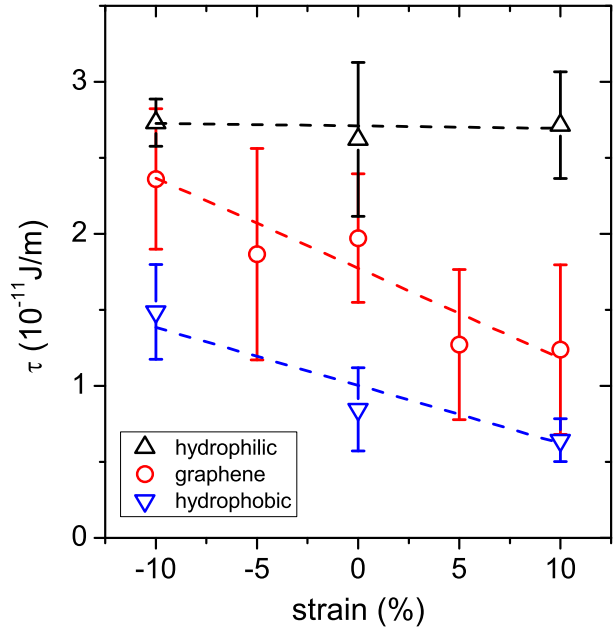}}
\caption{(color online) line tension fitted from Fig. \ref{fig2} using Eq. (\ref{eq1}) as a function of the applied strain on the substrate.}
\label{figlinetension}
\end{figure*}

Fitting the contact angle vs. droplet size to Eq. (\ref{eq1}) yields the values of line tension $\tau$ as well. The results are shown in Fig. \ref{figlinetension}. They are on the order of $(1.0-3.0) \times 10^{-11} $J/m, depending on the wetting properties of the substrates and applied strains (Fig. \ref{figlinetension}). The obtained magnitude of $\tau$ is consistent with the theoretical predictions and some recent experiments \cite{Pompe2000}. It is known that a substrate with a higher surface atom density and better wettability usually has a larger line tension $\tau$ \cite{Duncan1995, Amirfazli1998}. Stretching the graphene and the hydrophobic-surface will reduce the surface atom density and the wettability, thus leading to a dropped line tension value as observed in Fig. \ref{figlinetension}. In contrast, Fig. \ref{figlinetension} suggests that the hydrophilic-surface (a nearly constant line tension, $2.7 \times 10^{-11}$J/m) is almost independent on these two factors. It is worth noting that there is still a debate on the quantitative value of $\tau$ and data reported often have differences in several orders of magnitude \cite{Amirfazli2004, Weijs2011}.

In light of the inherent relation between the contact angle and the water-solid interfacial energy $\gamma_{SL}$, in the followings, we will report a linear relation between the $\cos\theta_{\infty}$ and the adsorption energy $E_{ads}$ of a single water molecule over a solid surface, which will help us understand the effects of strain on the $\cos\theta_{\infty}$ (Fig. \ref{fig2}). The Young-Dupre equation \cite{Israelachvili2011} correlated the macroscopic contact angle and the work of adhesion:
\begin{equation}
\label{eq2b}
\gamma _{LV} (1 + \cos \theta _\infty  ) = W/A_0
\end{equation}
where $W = (\gamma_{LV}+\gamma_{SV}-\gamma_{SL}) A_0$ is the work of adhesion, $A_0$ is the contact area, $\gamma_{SV}$, $\gamma_{SL}$ and $\gamma_{LV}$ are the interfacial free energy per unit area for the solid-vapor, solid-liquid, and liquid-vapor interfaces, respectively.

In our molecular model, the water-solid interaction energy is zero when they are separated infinitely far away from each other. Thus we can calculate the work of adhesion $W$ by summarizing the molecular pair-wise interactions between the water molecule and the solid surface atoms \cite{Israelachvili2011, Seemann2001, Mugele2012}, i.e.,
\begin{equation}
\label{eq2c}
W =  - \int_0^\infty  {E(h)\rho (h)A(h)} dh
\end{equation}
where $E(h)$ is the interaction energy of a single water molecule with a distance $h$ from the substrate, $\rho(h)$ are number density of the water molecules per unit area, and $A(h)$ is the cross section area of the water droplet. As usual, the summation of pair-wise vdw interactions between one water molecule and all carbon atoms on the monolayer surface can be replaced by an integration assuming the continuous distribution of carbon atoms \cite{ Israelachvili2011},
\begin{equation}
\label{eqE}
E(h) = \int\limits_0^\infty  {\frac{4}{{3\sqrt 3 a_{{\text{CC}}}^2 }}\phi \left( {\sqrt {r^2  + h^2 } } \right)dr}  = \frac{{16\pi \epsilon }}
{{3\sqrt 3 a_{{\text{CC}}}^2 }}\left[ {\frac{{\sigma ^{12} }}{{5h^{10} }} - \frac{{\sigma ^6 }}{{2h^4 }}} \right] = \left[ {\frac{{5\sigma ^4 }}
{{3h^4 }} - \frac{{2\sigma ^{10} }}{{3h^{10} }}} \right]E_{ads}
\end{equation}
where $E_{ads}$ is the lowest point of the potential well $E(h)$, which occurs at $h=\sigma$:
\begin{equation}
\label{eqEads}
   E_{ads}  = E(h = \sigma ) =  - \frac{{8\pi \epsilon }}{{5\sqrt 3 }}(\frac{\sigma }{{a_{{\rm{CC}}} }})^2  =  - \frac{{8\pi \epsilon }}{{5\sqrt 3 }}[\frac{\sigma }{{(1 + \varepsilon )a_{{\rm{CC}}_0 } }}]^2  \approx  - \frac{{8\pi \epsilon }}{{5\sqrt 3 }}(\frac{\sigma }{{a_{{\rm{CC}}_0 } }})^2 (1 - 2\varepsilon )
\end{equation}
The value of $E_{ads}$ represents the adsorption energy of single water molecule over a solid surface, which is linearly correlated with the applied mechanical strain $\varepsilon$ on the solid surface in Eq. (\ref{eqEads}).

Our previous MD study \cite{Xiong2011} showed that there is a depletion layer at the interface with a thickness of 0.2 nm, between 0.2 nm and 0.5 nm is the so called `first water layer' with a peak density $\rho$ about 2-3 times higher than the bulk value, between 0.5 nm and 1.0 nm is a second layer with density exhibit small oscillation around bulk water density, beyond 1.0 nm is the bulk water with a constant density $\rho_0$. This density profile $\rho(h)$ has a weak dependence on the strains applied on the surface. A similar density profiles are also observed for the hydrophilic and hydrophobic surfaces in this study. Thus, we can perform the integration [Eq. (\ref{eq2c})] in two separate parts:
\begin{eqnarray}
\label{eqlinear}
1 + \cos\theta _\infty   &=&  - \frac{1}{{\gamma _{LV} }}\int_{h_l }^{h_u }  \rho (h)E(h)dh - \frac{{\rho _0 }}{{\gamma _{LV} }A_0}\int_{h_u }^\infty   E(h) A(h) dh \nonumber \\
  &=&  - \frac{1}{{\gamma _{LV} }}\int_{h_l }^{h_u }  \rho (h)E(h)dh - \Delta _{tail}  \nonumber \\
  &=&  \left\{\frac{{-1}}{{\gamma _{LV} }} {\int_{h_l }^{h_u }  \rho (h)\left[ {\frac{{5\sigma ^4 }}{{3h^4 }} - \frac{{2\sigma ^{10} }}{{3h^{10} }}} \right]dh} \right\} E_{ads } - \Delta _{tail}
\end{eqnarray}
where the $h_l$ and $h_u$ represent the lower bound of the first water layer and the upper bound of the second, respectively.

In Eq. (\ref{eqlinear}), the first water layer has a monolayer thickness so that its cross section area is approximately the droplet contact area $A_0$. The rapid decay of $E(h)$ for $h > \sigma$ implies that the second term on the right side of Eq. (\ref{eqlinear}) is much smaller than the first one. Indeed, our calculations showed that $\Delta_{tail}$ is less than 5\% of the first term. Because the $\sigma$ is a constant (LJ potential between water and carbon atoms), the liquid-vapor interfacial $\gamma_{LV}$ is independent of the solid surfaces, and the change of water density $\rho(h)$ upon strain engineering has a small effect in the integral term, which we will discuss in detail later, we can conclude that cosine of the macroscopic contact angle $\theta_\infty$ should exhibit an approximate linear relation with respect to the adsorption energy $E_{ads}$.

\begin{figure*}[htbp]
\centerline{\includegraphics[width=0.6\textwidth]{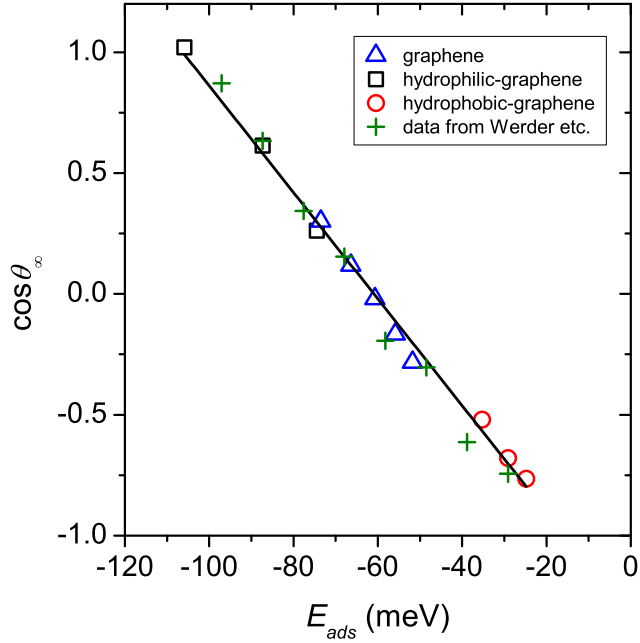}}
\caption{(color online) Cosine of the macroscopic contact angle $\theta_\infty$ as a function of the adsorption energy of a single water molecule on top of the strained graphene, hydrophilic-surface, hydrophobic-surface. We also included the simulated microscopic contact angle results of Werder etc. \cite{Werder2003}.}
\label{fig4}
\end{figure*}

Fig. \ref{fig4} summarizes the $\cos\theta_\infty$ directly obtained from our MD simulations (Fig. \ref{fig2}) as a function of $E_{ads}$ for the differently strained graphene, hydrophilic-surface and hydrophobic-surface. Indeed, it is a linear function, as predicted in Eq. (\ref{eqlinear}). The fitted value of the slope is $-22.0$/eV. The contact angle data from Werder etc. \cite{Werder2003} are also included in Fig. \ref{fig4} and the agreement is good. Since $E_{ads}$ is a linear function of the applied strain [Eq. (\ref{eqEads})], we can understand the linear relation between $\cos \theta_{\infty}$ and $\epsilon$ depicted in Fig. \ref{fig2}.

It is interesting to investigate the changes of water structures upon the strains of graphene layer and how such changes would determine the contact angle. In our previous publication \cite{Xiong2011}, we analyzed the water structure (density, radial density function and structure factor) at the interface of the strained `real' graphene. The two-dimensional radial distribution and structure factors of the first water layers on the $-10\%$, $0\%$ and $10\%$ strained graphene substrates are almost identical, while the water density profiles along the perpendicular direction of the substrates are similar except a $\pm 15\%$ change in the peak density of the first water layers. We have carried out analysis on MD simulations results on hydrophobic and hydrophilic surfaces. Similar conclusions are obtained. We found that the change of water density upon strain engineering has a small effect in the integral term in Eq. (\ref{eqlinear}). For `real' graphene, the integral term varies from $-22.2$/eV to $-20.7$/eV at a strain from $-10\%$ to $10\%$. For hydrophobic surface, it varies from $-18.9$/eV to $-17.0$/eV. And for hydrophilic surface, it is from $-22.5$/eV to $-22.2$/eV. All of them are close to the fitted slope $-22.0$/eV in Fig. \ref{fig4}. This shows our analytical model is consistent quite well with our MD results. This also indicates the change of $E_{ads}$ upon strain engineering plays a dominant role on cosine of the contact angle.

It is worth noting that in Ref. \cite{Werder2003}, Werder et.al. correlated the contact angle $\theta_{\infty}$ to the equilibrium adsorption energy, in which they concluded that only in a certain range, the contact angle follows a linear relationship with the $E_{ads}$. We believe that Eq. (\ref{eqlinear}) (Fig. \ref{fig4}) is a better model to understand the relation between the surface wettability and the chemical/physical interactions of liquid and solid surfaces.

Since our model is derived without presumption on the atomic structures of the substrate, it can be easily applied to 3-dimensional substrates (e.g. graphite) by summing the interaction energy between water molecules and the different atomic layers of substrate, although the surface atomic density of a substrate and the averaged $E_{ads}$ might not necessary be linear functions of the applied strain/stress. It is worth noting that since the Lennard-Jones interaction decays rapidly over distances, the interfacial energy is mainly determined by the first atomic layer of substrates \cite{Rafiee2010, Shih2012}. We believe our results can provide some valid indications for a broad class of three dimensional substrates.

Strain up to  30\% can be readily exerted on graphene in experiments \cite{Ni20082, Mohiuddin2009, Kim2009}. For example, an epitaxial strain of $\sim \pm1\%$ builds up in the graphene when it is grown on different substrates \cite{Kim2009, Pan2009, Ni2008, Wintterlin2009}; uniaxial strain up to $\sim 1.3\%$ to a graphene monolayer can be applied by using two- and four-point bending setups \cite{Mohiuddin2009}; and uniaxial strain ranging from 0 to $30\%$ can be achieved for large-scale graphene films transferred to a pre-stretched substrate \cite{Kim2009}. There are also many ways to reversibly control the strains applied on substrate materials by mechanical stresses or electric fields. For example, reversible strain obtained in Fe-Pd single crystals by compressive stress-induced martensite variant rearrangement is reported to exhibit as high as 5\% \cite{Yamamoto2007}; Zhang etc. obtained a large reversible electric-field-induced strain of over 5\% in BiFeO3 films \cite{Zhang2001}. We, therefore, believe strain engineering can be a promising way to reversely control the surface wettability in practice. In practice, compressive strain often leads to ripples or folding of graphene. In this manuscript, the study on the compressive strain serves a theoretical interest. We aim to obtain a more complete picture of the relation between contact angle and molecular adsorption energy and substrate strain.

Strain engineering, in our previous study, has been predicted to significantly change the slip length of water over a graphene layer \cite{Xiong2011}. It is thus interesting to investigate the correlation between the wettability and the slip length. It is often believed that a hydrophobic surface often has a larger slip length than that of a hydrophilic surface \cite{Voronov2008}. Strain engineering the wettability and the slip length turn to be an exceptional case. In Fig. \ref{fig2}, with a strain from $-10\%$ to $10\%$, the graphene is becoming more hydrophobic (e.g., contact angle increasing from 72.5$^\circ$ to 106$^\circ$). The slip length obtained in our previous MD simulations was reduced from 175 nm to 25 nm. Liquid water can slip on a hydrophilic surface, even with a larger slip length than that on a hydrophobic surface \cite{Ho2011}. To understand such a counter-intuitive observation, we should note that Eq. (\ref{eqlinear}) implies the dependence of contact angle on $E_{ads}$, which is an average of the van der Waals potentials over the whole solid surface (at $h=\sigma$), whereas it is the energy barrier (i.e., the corrugation of the vdw potential profile) experienced by water molecules over a solid surface who defines the slip length \cite{Xiong2011}. Stretching a graphene layer results in a smaller $E_{ads}$ [Eq. (\ref{eqEads})] and thus a higher contact angle $\theta_{\infty}$, but it leads to an enhanced energy barrier and thus a reduced slip length \cite{Xiong2011}. This insight may help understanding the observed controversial correlations between the contact angle and the slip length in experiments \cite{Voronov2008, Neto2005}.

\section{Conclusions}

To summarize, using graphene sheet as a sample material, our MD simulations show that the wettability of a solid surface can be controlled by mechanical strains. Overall, the cosines of the contact angles exhibit a linear relation with respect to applied strains. For a graphene surface, the contact angles can be tuned from 72.5$^\circ$ to 106$^\circ$ under biaxial strains ranging from $-10\%$ to $10\%$. For droplets on intrinsic hydrophilic surfaces (at zero strain), the variation of the contact angle is more sensitive than that of a hydrophobic surface. To understand the strain engineering effect, we developed an analytical model to reveal a linear relationship between the $\cos\theta_\infty$ and the adsorption energy $E_{ads}$ of a single water molecule over the substrate surface. The applied mechanical strains change the $E_{ads}$ and consequently alter the contact angle. We believe that the linear relationship between the cosine of the contact angle and the adsorption energy $E_{ads}$ is a general model to describe the surface wettability and the chemical/physical interactions of liquids and solid surfaces. Thanks to the ease of reversibly applying mechanical strains in MEMS/NEMS, we propose that strain engineering can serve as an effective means to achieve the reversibly control of surface wettability.

\begin{acknowledgements}

Q.S.Z. acknowledges the financial support from the National Science Foundation of China (NSFC) through Grant No. 11172149, the IBM World Community Grid project "Computing for Clean Water", and the Boeing-Tsinghua Joint Research Project "New Air Filtration Materials". J.Z.L. acknowledges seed grant 2012 from engineering faculty of Monash University. This work was supported by an award under the Merit Allocation Scheme on the Australia NCI National Facility at the ANU.

\end{acknowledgements}

\providecommand*{\mcitethebibliography}{\thebibliography}
\csname @ifundefined\endcsname{endmcitethebibliography}
{\let\endmcitethebibliography\endthebibliography}{}

\end{document}